\documentclass[manuscript,screen,nonacm]{acmart}

\AtBeginDocument{%
  }

\setcopyright{none}
\usepackage{xspace}
\usepackage{multirow}
\usepackage{booktabs}
\usepackage{CJKutf8}
\usepackage{xcolor}
\usepackage{needspace}

\newcommand{\systemname}{GELLI\xspace}

\begin{document}

\title{GELLI: A Generator for Experiential Language Learning Interactions}

\author{Danilo Symonette}
\affiliation{\institution{Stanford University}\country{USA}}
\author{Alan Y. Cheng}
\affiliation{\institution{Stanford University}\country{USA}}
\author{Taiga Ikedo}
\affiliation{\institution{Stanford University}\country{USA}}
\author{Jose Manuel Rosales Juarez}
\affiliation{\institution{Pontificia Universidad Católica del Peru}\city{Lima}\country{Peru}}
\author{Roy Pea}
\affiliation{\institution{Stanford University}\country{USA}}
\author{Parastoo Abtahi}
\affiliation{\institution{Princeton University}\country{USA}}
\author{Jane L. E}
\affiliation{\institution{Stanford University}\country{USA}}
\affiliation{\institution{National University of Singapore}\country{Singapore}}
\author{James Landay}
\affiliation{\institution{Stanford University}\country{USA}}
\renewcommand{\shortauthors}{Symonette et al.}

\begin{teaserfigure}
  \includegraphics[width=\textwidth]{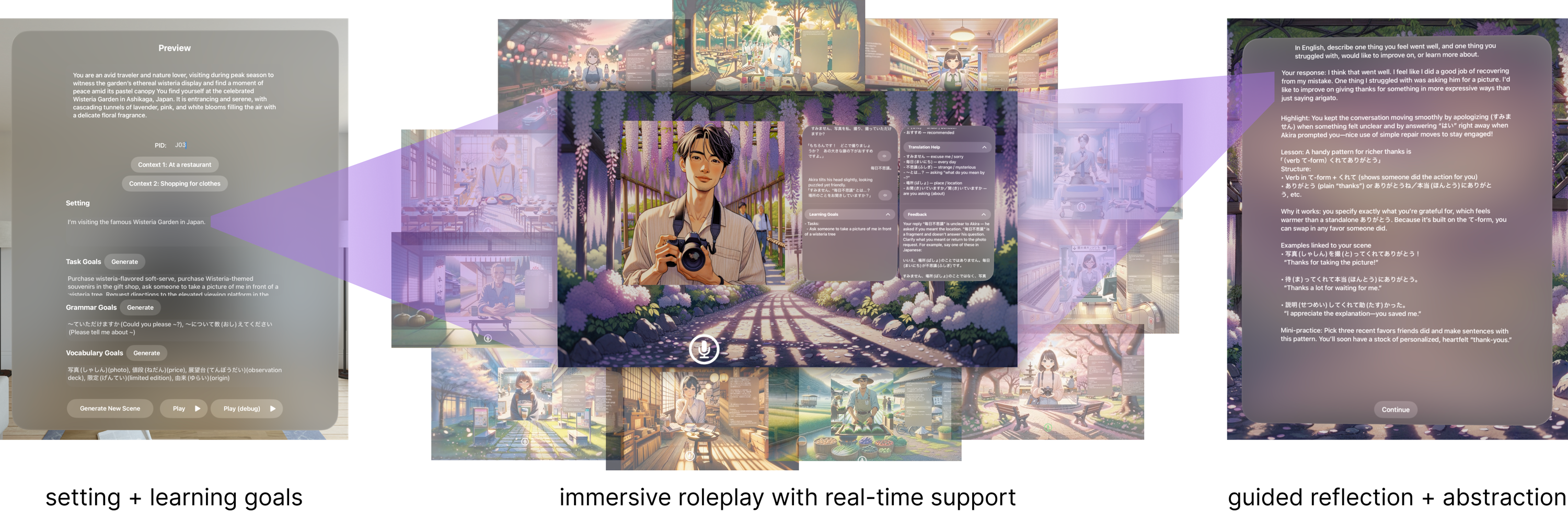}
  \caption{We present \systemname, a system for generating narrative practice worlds from the learner's 
own goals. The learner specifies a setting and learning goals 
(left), and \systemname generates situated scenes with a character 
to converse with, supported by translations, suggestions, and 
corrective feedback (middle). After each scene, the learner 
reflects and receives feedback before entering the 
next scene (right). }
  \label{fig:teaser}
\end{teaserfigure}

\begin{abstract}

Learning a second language requires extensive practice across situations personal to each learner, yet the most effective existing approaches --- narrative situations designed around specific learning goals and delivered through media --- demand substantial authoring effort, limiting practice to the scope of what was built. We present GELLI, a system that generates narrative practice worlds from a learner's goals: situated scenes with characters to converse with, support alongside the interaction, and reflection, all produced at runtime with no pre authored content. In a within-subjects study with 40 learners of Japanese, GELLI produced greater vocabulary recall and confidence than expert-authored videos targeting the same goals, with no significant difference in speaking performance. Learners turned remembered or imagined situations into experiences they could shape and learn through. Support made experimentation feel safe, while unexpected narrative developments and language suggestions gave them new possibilities to pursue beyond what they initially intended to practice.
 
\end{abstract}

\keywords{language learning, generative AI, experiential learning, narrative, mixed reality, conversational practice}

\maketitle
\hypersetup{pdfauthor={Danilo Symonette, Alan Y. Cheng, Taiga Ikedo, Jose Manuel Rosales Juarez, Roy Pea, Parastoo Abtahi, Jane L. E, James Landay}}

\section{Introduction}

Building proficiency in a second language can take hundreds to thousands of hours of sustained practice in a variety of situations~\cite{StateDeptForeignLanguageTraining, LanguageTestingInternational}. Practice grounded in lived experience is especially valuable, as personally meaningful activity in contextually-aligned language-learning environments supports comprehension, transfer, and long-term participation~\cite{priniski2018making, nachtigall2022stirring}. However, relevant real life situations, in addition to often being unavailable, can also be too chaotic and intimidating for much learning progress ~\cite{kinginger2009sa,krashen1982principles}. Thus, a core challenge for language learning technology is to deliver experiences that feel situated in relevant social and physical contexts while also providing the support, structure, and accessibility that authentic situations often cannot.

Prior work on creating this type of practice has largely focused on designing narrative situations around specific learning goals and delivering them through media ~\cite{rowe2011ijaied,lester2012narrative,calvert2022ajet,oakstory}. Narrative provides context that makes actions meaningful and helps retention~\cite{bruner1990acts,schank1977scripts,lee2023factors}, the media establishes a sense of place and social presence that deepens engagement~\cite{reeves1996media,bruni2022digital,godwin2023presence}, and alignment with learning goals ensures that the experience is not only engaging, but also pedagogically productive~\cite{rowe2011ijaied,oakstory}. However, creating these experiences requires substantial authoring effort, limiting practice to the predetermined situations and learning goals of the author's design~\cite{jemmali2018EducationalGameDesign, lester2012narrative, calvert2022ajet}. What remains missing in previous work is a way to provide such practice for whatever goals a learner brings.

Apart from the effort of producing the representational media, much of the authoring burden in narrative-centered learning comes from the top-down coordination of plot with pedagogy. Every scene must be scripted to advance both the story and the learning goals, and the learner's actions must be anticipated in advance ~\cite{rowe2011ijaied,lester2012narrative}. Emergent narrative~\cite{aylett1999aaaiFS} offers a way around this constraint, through narrative designed to arise from the learner's interactions rather than from a scripted plot. But without a plot to organize the experience, something else must provide the necessary pedagogical structure. Experiential learning theory~\cite{Kolb1984} suggests what that structure should be: a recurring cycle of experimentation, reflection, and abstraction around instances of concrete experience. We envision a system that generates narrative situations to provide this concrete experience, constructs them around successive subsets of the learner's goals, and supports the learner through the full cycle without preauthored content. 

Realizing this vision involves overcoming significant technical and design challenges. Narrative must coherently emerge without authored plot, engagement must arise without authored environments, and pedagogical structure must be achieved without constraining the learner's actions. To navigate these challenges, we design and develop \systemname: a Generator for Experiential Language Learning Interactions. Given a set of learning goals, \systemname selects a subset for each scene and generates a situation that makes the communicative actions those goals require natural and necessary, featuring a character for the learner to converse with. To provide immersion, \systemname generates 2D scene and character imagery, pairs them with voiced narration and dialogue, and delivers them at room scale through a mixed reality headset. Support windows help the learner act without interrupting the roleplay and provide corrective feedback. After each scene concludes, the system prompts the learner to reflect on what happened and generates feedback grounded in their reflection and the conversation transcript, then uses remaining goals to produce the next scene. In this way, \systemname situates the learner in a generated narrative world that both responds to their actions and shapes what they practice, structured around the experiential learning cycle.

To examine whether runtime-generated practice could support learning despite risks of incoherence, limited engagement, and goal misalignment, we evaluated \systemname with 43 learners of Japanese, including a controlled within-subjects comparison with 40 participants. We compared \systemname with expert-authored conversation practice videos targeting the same learning goals. These videos provided a reference for pedagogical alignment: every target goal appeared in context through voiced characters, illustrated scenes, and bilingual subtitles, delivered on the same mixed reality headset.

Learners recalled significantly more target vocabulary and reported significantly greater confidence after \systemname than after the videos. Speaking performance improved in both conditions, with no significant between-condition difference. Coding of session logs showed that generated scenes created opportunities for most assigned goals and that learners attempted most available opportunities. Participants also reported significantly greater narrative presence, social presence, spatial presence, and emotional engagement than with the videos. Their accounts linked immersion to believable interaction, imagination, and consistency among generated elements and their expectations. Learners valued authoring situations for preparation and play and described the unfolding narrative as something they helped create, while identifying moments when steering toward learning goals interfered with the direction they wanted to pursue.

We contribute:

\begin{itemize}
    \item (1) empirical evidence that runtime-generated narrative scenes can support learning and create opportunities for targeted language practice while preserving conversational agency;
    \item (2) an account of how learners experience narrative immersion, authoring, and steering in generated worlds without a pre-authored plot;
    \item (3) design implications for generated practice covering steering toward learning goals, consistency across representations, and support for authoring and exploration.
\end{itemize}

\section{Related Work}

Below, we survey contributions from adjacent fields that ground our interdisciplinary approach. 

\subsection{Interactive, Adaptive, and AI-Driven Learning Systems}

Intelligent tutoring systems demonstrate that learner 
modeling, adaptive sequencing, and timely feedback improve 
learning~\cite{vanlehn2006behavior,ma2014its}. Large 
language models have since expanded what can be generated 
for learners. Generative systems can now produce educational content in an unprecedented scope, from stories that embed target vocabulary~\cite{peng2023storyfier} to practice tasks and environments generated procedurally~\cite{rodrigues2019procedurally, Wang2023_VRPCGedu}. In language learning specifically, LLM-based roleplay tools such as Speak, ELSA AI, and TalkPal have made interactive conversational practice widely accessible and increasingly incorporate scenario drafting, task tracking, and post-conversation feedback~\cite{speak_live_roleplays,elsa_ai_website,talkpal_roleplay_website}. Research systems like ChitterChatter align AI-generated conversation to instructor-defined curriculum goals~\cite{geathers2025chitterchatter}, while Rehearsal demonstrates targeted practice with feedback for conflict resolution scenarios~\cite{ShaikhEtAl2023Rehearsal}. However, work on AI roleplay for language learning also surfaces persistent challenges: LLM dialog can drift from objectives, break character, or fail to create opportunities for target forms~\cite{choi2024examining,glandorf2025grammar}. Most of these systems still treat the interaction primarily as a single conversation with an agent, without the cognitive and emotional benefits of narrative design, or the reliable pedagogical alignment that comes from designing scenarios around specific learning goals.

\subsection{Narrative-based Learning Systems}  %

Narrative-based learning uses storytelling structures to provide context for action and meaning. Stories provide schemas, scripts, and causal links that aid comprehension and retention~\cite{bruner1990acts,schank1977scripts,vandijk1978psychrev}, while narrative transportation increases engagement and supports attitude change~\cite{green2000transportation,busselle2009measuring}.

Interactive narrative environments organize learning tasks within storyworlds. Systems such as Crystal Island tie learning supports to plot progression~\cite{rowe2009crystal}, while larger worlds like Quest Atlantis use story goals and 
consequential choices to structure inquiry~\cite{barab2005quest}. 
A key challenge in these systems is maintaining pedagogical alignment when learners deviate from the intended path. 
Research in interactive digital narrative addresses this 
issue through drama management and narrative 
planning~\cite{riedl2013interactive}, in which an experience 
manager coordinates narrative goals with pedagogical goals to 
produce a coherent experience that is both engaging and 
educationally productive. Recent work such as Oak 
Story~\cite{oakstory} extends this approach by using LLMs to 
give learners greater agency within the narrative while still driving them toward pre-authored narrative events and learning activities. Still, these approaches manage pre-authored assets and organize scripted or pre-determined narrative events that the learner must progress through.

Emergent narrative approaches offer an alternative to plot-driven coordination. Rather than scripting narratives top down, these systems allow stories to arise from autonomous character interaction, as demonstrated in educational drama systems like FearNot!~\cite{aylett2005fearnot}. This approach trades in the coherence guaranties of scripted plots for greater learner agency and spontaneity, a tradeoff we revisit in Section~\ref{sec:design}.

\systemname builds on this tradition but operates under a different constraint: rather than coordinating pre-authored narrative and pedagogical events, it must generate both at runtime from the learner's goals. This shifts the design challenge from narrative management of authored content to narrative generation of goal- conditioned content, which we address through emergent narrative and experiential learning theory.

\subsection{Immersive Learning Environments}

VR and XR research has shown that immersive environments can enhance educational experiences through presence and embodiment, as well as support situated learning~\cite{dede2009science,lan2020vr,lave1991situated}. Foundational research shows that people apply social and physical rules to digital media as if they were real, treating mediated interactions as genuine encounters~\cite{reeves1996media}. This helps explain why simulated experiences can evoke emotions and learning outcomes similar to lived events.

Within language learning, XR-assisted practice is associated with improved oral proficiency, greater motivation, and lower anxiety~\cite{lan2020vr,Chen2022XRFrontiers,Chen2022VRSust,Yu2024FLA,berns2021review}. However, the landscape is narrower than appears. Review articles in the literature note that few XR language learning applications explore the real potential of VR for novel, immersive, or experiential learning approaches~\cite{godwin2023presence, berns2021review}. Most operate as highly immersive flashcard systems~\cite{legault2019immersive} or present fixed tourist-like scenarios with limited scope and interactivity~\cite{godwin2023presence}.

Research on `presence' further complicates the picture. Although stronger feelings of presence often enhance engagement, the evidence that full 3D immersion improves learning outcomes over lighter alternatives is mixed~\cite{makransky2022benefits, hamilton2021immersive}. Factors such as modality, previous experience, and cognitive load introduce nuance, and richer environments can decrease learning when the medium overwhelms the message \cite{ahn2022unintended,makransky2019adding}.

\section{Design Tensions and Goals}\label{sec:design}

In many authored approaches, experts embed their pedagogical and narrative expertise into a unified design: they craft narrative that organizes and contextualizes interaction, develop media that helps learners feel present in an imagined social situation, and script dialogue aligned with specific learning goals. This process produces effective practice, but it is complex and costly, limiting practice to the scope of what was built and to the interactions anticipated by the designer~\cite{lester2012narrative}. To bring this kind of practice into line with whatever goals the learner brings, with the learner free to say and do what they choose, the narrative, media, dialog, and pedagogical structure must all be generated at runtime. This demand introduces design tensions to be resolved: how to organize narrative, achieve presence, and structure learning when each must be generated rather than designed by hand. In the following, we discuss these tensions and the theoretical lenses we used to navigate them, deriving design goals that guided our implementation decisions.

To ground these decisions in domain expertise and learner experience, we iteratively prototyped with two Japanese language instructors and eight learners of Japanese. Three principles from this process informed our design: support should preserve productive struggle, support should remain accessible without interrupting the roleplay, and learners need control over goal specification and pacing. 

\subsection{Generating Narrative without Authored Plot}

\subsubsection{Tension.} In authored narrative-centered learning environments, plot serves as both the narrative and pedagogical backbone: it determines what the learner encounters, in what order, and with what dramatic arc \cite{rowe2011ijaied,lester2012narrative}. The coordination of plot with pedagogy is what makes these systems effective, but it is also what makes each scenario expensive to author and rigid to adapt.

\subsubsection{Theoretical lens.} Emergent narrative~\cite{aylett1999aaaiFS} points to an alternative: rather than scripting a plot, the system could generate social situations and let narrative arise from the learner's interactions with characters who behave according to their roles, removing the need to author every path and giving the learner genuine conversational agency. However, without a plot organizing the experience toward specific objectives, the interaction could be engaging but pedagogically aimless. In a dynamically generated narrative, learning goals could serve as the organizing unit in place of plot, with the scene as the atomic unit of both narrative and pedagogy.

\paragraph{\textbf{DG1: Generate narrative scenes around subsets of learning goals.}} A scene should be a single situated interaction between the learner and a generated character, organized around a subset of the learner's goals that cohere within one social encounter. Within the scene, the character should stay in role and create opportunities for the learner to practice target forms, while responding to unexpected input as someone in that social role would, rather than breaking character to enforce topic adherence. Pedagogical relevance should be maintained through the social logic of the situation, not through scripted constraints on what the learner can say.

\paragraph{\textbf{DG2: Produce narrative continuity from goal progression and conversation history.}} Successive scenes should be shaped by the remaining goals and the consequences of what has already occurred, producing narrative continuity without authored plot structure. For example, a learner whose goals include checking for items at a lost-and-found venue and taking the train to a destination might first find themselves at the lost-and-found, explaining to an attendant that they lost their wallet. In the next scene, the system might generate a situation where they must ask a stranger to borrow money for a new transit card (narrative consequence), in order to take the train (remaining learning goal). 

\subsection{Representing the Narrative World}

\subsubsection{Tension.} Authored approaches achieve a sense of presence and engagement through a wide range of media, and the evidence on which medium best supports learning is mixed~\cite{hamilton2021immersive,makransky2019adding,makransky2022benefits}. For runtime generation, we face three key constraints that shape our selection of representational strategy. First, generating richer assets increases latency, cost, and the risk of visual incoherence. Second, richer environments can be distracting and increase cognitive load without improving learning. Third, fully-rendered environments reduce narrative flexibility since every narrated event must be consistent with what is shown, constraining narrative to existing visuals or requiring mid-conversation regeneration. Lighter representations leave room for narration and imagination to fill the gaps, making the generated narrative more resilient to conversational contingencies.

\subsubsection{Theoretical lens.} Research on narratives in extended reality distinguishes between system immersion, the technological envelopment of the user's senses, and narrative immersion, the psychological engagement produced by imagery, characters, and storytelling~\cite{bruni2022digital}. A sense of place and social connection can be evoked largely independently of environmental fidelity~\cite{bruni2022digital, ryan2001}, as spatial immersion applies across narrative media and emotional engagement is driven by characters and events rather than explorable environments. This motivates a representational strategy in which imagery, narration, and responsive characters may be sufficient to produce presence and engagement for runtime generation.

\paragraph{\textbf{DG3: Target spatial and emotional immersion through generated imagery, narration, and voice.}} Representation should maximize narrative flexibility while minimizing latency. In addition, it should establish a sense of place through visual context, a sense of social presence through a visible voiced character, and bridge representational gaps through narration. 

\subsection{Structuring Learning Under Learner Agency}

\subsubsection{Tension.} A third tension arises from the need to structure learning within an interaction in which the learner has full conversational agency. In authored approaches, designers can calibrate exactly when and how to introduce support, feedback, and reflection because they know what the learner will encounter. Under full generativity, the system cannot predict what will happen, so pedagogical structure must emerge from the interaction itself.

\subsubsection{Theoretical lens.} Experiential learning theory~\cite{Kolb1984} provides the framework we use to organize the explicit pedagogical structure on top of the situational context. Kolb describes learning as a recurring cycle of concrete experience, reflective observation, abstract conceptualization, and active experimentation. We treat each generated scene as a concrete experience and organize support around the remaining phases of the cycle.

\paragraph{\textbf{DG4: Align support with experiential learning cycle.}} In-scene support should facilitate active experimentation and abstract conceptualization during the concrete experience, while between scene reflection should facilitate reflective observation and connect the learner's experience to broader principles.

\paragraph{\textbf{DG5: Preserve learner agency and productive struggle.}} The learner should control what they say, when they speak, and when they move on. Support should preserve learner agency in producing their own responses. Support should not interrupt the roleplay. We derived this goal in part from our iterative design process with teachers and learners, described above. 

\section{\systemname: A Generator for Experiential Language Learning Interactions}\label{sec:system}

\begin{figure*}[t]
    \centering
    \includegraphics[width=\textwidth]{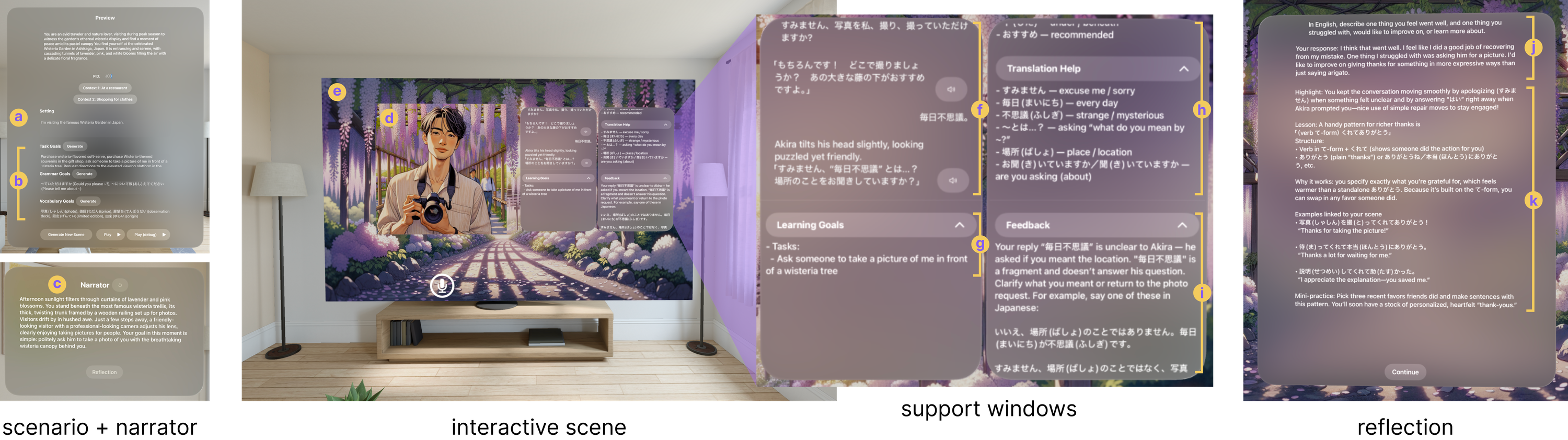}
    \caption{The interface of \systemname. The scenario design screen lets learners specify a setting (a) and learning goals (b). The narrator introduces the generated scene (c). During interaction, a foreground character (d) and background image (e) represent the situation. Support windows show the conversation transcript (f), learning goals (g), response suggestions and translation help (h), and corrective feedback (i). After the scene, learners reflect on their experience (j) and receive generated feedback (k).}
    \label{fig:interface}
\end{figure*}

\begin{figure*}
    \centering
    \includegraphics[width=\textwidth]{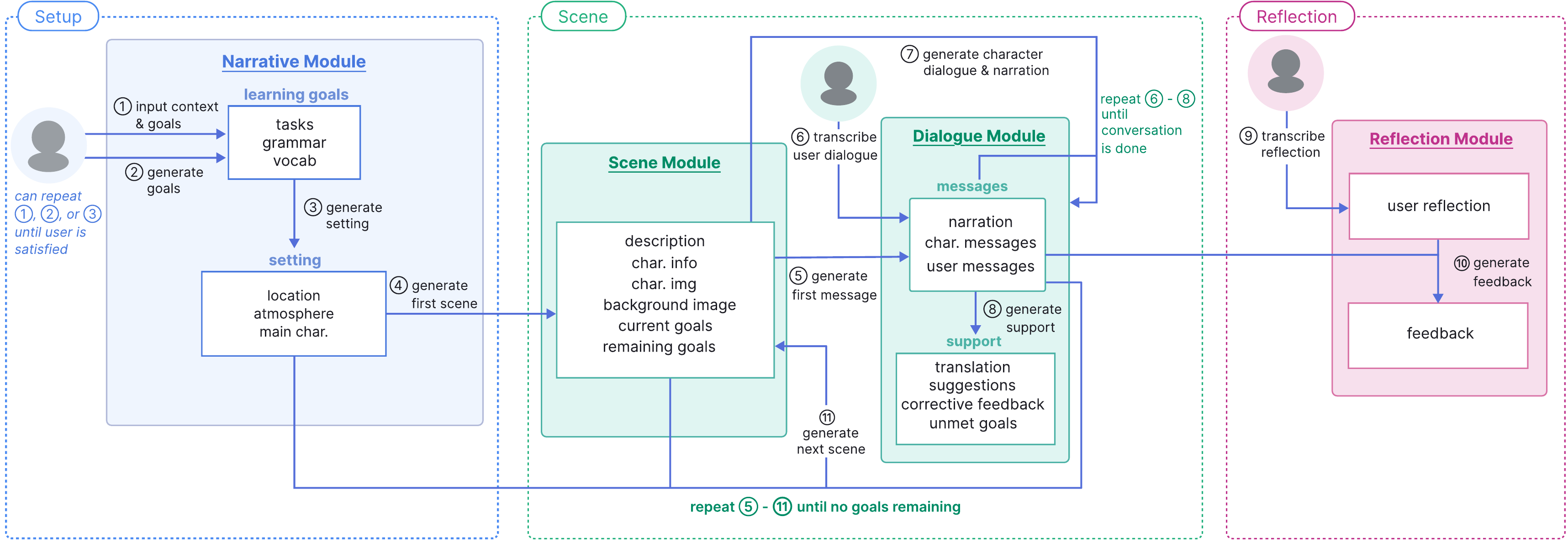}
    \caption{A system diagram showing the flow of data through a session in \systemname. First, the learner inputs the context and any learning goals they want to practice (1), generating any goals they did not specify (2). The system then generates the setting (3) and sets up the first scene (4). In the scene, the system generates the first message from the AI character (5). Then, the learner speaks (6), the system generates character dialogue and narration in response (7), and it generates support text (8). These steps repeat until the conversation reaches an end, at which point the learner records a reflection (9) and receives generated feedback (10). Finally, the system generates the next scene (11) and the cycle repeats until all learning goals have been covered.}
    \label{fig:dataflow}
\end{figure*}

\systemname implements the design goals described above through a coordinated multi-module architecture (see Figure~\ref{fig:interface} for the interface and Figure~\ref{fig:dataflow} for the system data flow). The learner specifies a setting and learning goals, then enters a series of interactive scenes where they converse with a generated character, receive off stage support, and complete a reflection activity after each scene. In the following, we describe each stage of the interaction and the architecture that coordinates them, as well as the design goal (DG) of Section~\ref{sec:design} which informed its design.

\subsection{Goal Specification}

The session begins with a menu where the learner specifies what to practice. Separate input fields accept a setting description, task goals, grammar goals, and vocabulary goals. The learner can fill in any subset of these and have the system generate the rest: a learner who specifies only a setting receives AI-generated tasks, grammar, and vocabulary aligned to that context. Learners can also generate goals first and have the system suggest a setting. Once at least one learning goal is present, the learner generates a preview describing their role and the scenario, regenerating until satisfied before beginning. This design allows learners who know exactly what they want to practice to specify it precisely, while learners who are uncertain can start from a setting and let the system suggest what to work on.

\subsection{Scene Generation and Interaction}

A narrative module selects a subset of learning goals for each scene, choosing goals that cohere within a single social encounter (DG1). Given the setting, atmosphere, character description, and remaining goals, it generates a scene description, a character with a role appropriate to the situation, and the scene's specific goal set. The prompt instructs the module to choose goals ``that the user can achieve by speaking or listening to the character in the scene'' and to scope each scene to what can be achieved in one static setting. The module also maintains a record of the remaining goals, removing the goals assigned to each scene when that scene is generated.

A scene module then generates a landscape-orientation background image and a foreground character image from the same detailed scene description, both displayed at room scale in the learner's physical space through a mixed reality headset (DG3). A narrated scene description is read aloud, orienting the learner to the situation, the character, and the objectives of the scene (DG3). 

The learner converses with the character using push-to-talk speech input (DG5), then their speech is transcribed and added to a message log. A dialog module generates the character's response and accompanying narration based on the full message log, the current scene's goals, and the scene description. The character's prompt instructs it to stay in role, keep the scene's learning goals in mind, create opportunities for the learner to practice them, and to respond with in-role confusion when the learner makes errors, triggering the support system rather than coaching directly (DG5). Narration describes the character's behavior in third person and in English, bridging what the learner sees with what is happening and helping the learner follow the social dynamics even when they cannot fully parse the character's Japanese dialog.

When the conversation has reached a natural close, the dialog agent sends a system message indicating the conversation has ended, but the learner may decide whether to continue the conversation or proceed to reflection (DG5).

\subsection{Support Windows}

To ensure that learners have the guidance and structure necessary to learn from the experience, \systemname provides support information in panels positioned adjacent to the character image, accessible without leaving the roleplay (DG5). \systemname generates the following support from the most recent messages:

\paragraph{\textbf{Dialogue transcription.}} The full conversation between the character and the learner. A small replay button accompanies any Japanese text so that the learner can hear it again.

\paragraph{\textbf{Learning goals.}} The list of learning goals assigned to the current scene. This list is updated as the learner exercises them, helping the learner to stay oriented to what they are practicing.

\paragraph{\textbf{Translations.}} Key words and phrases from the character's most recent dialog, translated for the learner with furigana for kanji.

\paragraph{\textbf{Response suggestions.}} Vocabulary and short phrases that the learner could use in responding to the character. These are never full sentences, preserving the need for the learner to construct their own utterances (DG5).

\paragraph{\textbf{Real-time corrective feedback.}} When the learner's response does not make sense in context or shows misunderstanding, the character responds with in- role confusion, asking them to repeat or clarify. In this case, the response suggestions are replaced by corrective feedback that tells the learner what they may have done wrong, provides a model correction, and offers a metalinguistic explanation of the underlying rule.

These supports are generated by a dedicated support agent that monitors the conversation independently of the dialog agent. It has access to the full message log, the scene's learning goals, and the overall session goals, allowing it to tailor support to both the immediate conversational context and the broader learning objectives.

In addition to addressing learner needs, these supports were designed to facilitate two phases of the experiential learning cycle (DG4). Translations and response suggestions give learners a concrete alternative to try on their next turn, creating immediate opportunities for active experimentation (AE). When real time feedback addresses a mistake by explaining a general rule and how or when to use it, it supplies an opportunity for abstract conceptualization (AC), and the interface then invites an immediate retry, linking AC and AE in the same exchange. In cases where feedback is corrective without a stated rule, it still advances AE by indicating a specific fix to attempt on the spot.

\subsection{Reflection and Feedback}

After each scene, the learner answers a brief spoken prompt: ``Describe one thing you feel went well and one thing you struggled with or would like to improve on.'' Their response is transcribed.

This reflection is designed to facilitate Reflective Observation (RO) by asking the learner to articulate successes and difficulties in relation to the just completed interaction (DG4). Using that self report together with the scene transcript and learning goals, a reflection module then generates feedback in three parts:

\paragraph{\textbf{Highlight.}} Reinforces a specific success from the conversation, with one additional observation of something the learner did well.

\paragraph{\textbf{Lesson.}} Provides a focused generalization, such as a grammar rule, a grammatical pattern, or a discourse strategy, aligned with the learner's stated struggle. This facilitates Abstract Conceptualization (AC).

\paragraph{\textbf{Review of learning goals.}} Describes how well the learner met each goal assigned to the scene and what they could do differently for any that were missed.

Together, the prompted self reflection and generated feedback facilitate the progression through RO and AC phases, setting up opportunities for AE in subsequent scenes, where the learner may encounter situations that call for applying the articulated takeaway.

After reflection, the narrative module receives the conversation history from the previous scene along with its setting, atmosphere, and character descriptions. Its prompt instructs it to ``maintain narrative coherence'' by taking this information into account when selecting the next subset of goals and generating the next scene description (DG2). As each scene is generated, the list of remaining goals is updated, and this loop continues until all learning goals have been covered or the learner ends the session.

\subsection{Additional Implementation Details}

\systemname coordinates these functions through a multi-agent architecture where separate agents handle narrative, dialog, support, goal tracking, and reflection, each grounded in the same scene context and learning 
goals but only accessing information necessary for its role (e.g. the dialog agent does not see support messages). Cross-scene coherence is maintained by passing the previous scene's conversation history, situation, and character details to the narrative agent when generating each successive scene. The system is built in SwiftUI on Apple Vision Pro using OpenAI models (o3 for text generation, dall-e-3 for imagery, gpt-4o-mini-tts for voice and whisper-1 for transcription), delivering generated imagery at room scale with spatial audio. We treat the headset as an implementation choice and do not isolate its effects in our evaluation.

\section{Study Design}

The goal of \systemname is to produce practice that is reliably aligned with learning goals, situated, and engaging ---  properties that authored approaches achieve through extensive design --- while allowing learners to practice whatever goals they choose with full agency and no pre-authored content. Our approach is to generate such practice at runtime, structured around the experiential learning cycle. This `high wire' approach introduces significant risks: generated scenes and imagery may be incoherent or distracting, the lightweight representation may not produce the sense of place and social presence it is designed for, the conversation may drift from learning goals, and the generated support and feedback may not be pedagogically adequate.

To evaluate the degree to which \systemname achieves its goal despite these risks, we compare it with expert-authored conversation practice videos\footnote{Cafe video: \url{https://www.youtube.com/watch?v=1oAOTbPzrg4}; shopping video: \url{https://www.youtube.com/watch?v=gfZoqjYk2Vw}.} that aim the same learning goals in a within-subjects study. The videos represent the pattern described in Section 2: narrative situations designed around specific learning goals, provisioned through media. We selected conversation practice videos because these were cited by Japanese conversation class instructors as representative of what they recommend to their students for out-of-class practice. An author with instructional design expertise and Japanese fluency analyzed each video and cross referenced its content with a Japanese language textbook to derive a precise set of tasks, grammar patterns, and vocabulary. Then these learning goals were provided as input to \systemname. During the video condition, every participant saw identical, carefully sequenced practice; during the  \systemname condition, every participant experienced a unique generated scenario. No two \systemname sessions were identical.

We investigate the following research questions:

\begin{itemize}
    \item \textbf{RQ1: Did they learn?} Given the agency to say anything and the openness of LLM generation, do learners still learn from generated roleplay practice compared to expert-authored practice 
    targeting the same goals?
    \item \textbf{RQ2: Did they feel immersed in a narrative world?} Can narrative generated on the fly produce narrative immersion without plot? What forms does it take, what does it depend on, and what did the learners say it did for them?
    \item \textbf{RQ3: How did they experience the co-authoring of the narrative with the AI?} How do learners engage with narrative authoring, steering, and continuity, jointly shaped by them and AI?
\end{itemize}
 
\subsection{Participants}

We recruited 47 participants through flyers, mailing lists, and word of mouth. Eligibility requirements included being at least 18 years old, the ability to come to our institution in person, at least one year of university-level Japanese language study or equivalent, and the ability to read text comfortably at a medium distance without glasses (as the Apple Vision Pro is not compatible with glasses).

Two participants were pilots and are excluded from analysis. Three did not complete the full study due to time constraints, and two have missing data due to collection errors. Three of these five were able to use the tool and provide feedback, yielding $N = 43$ for qualitative analysis and $N = 40$ for quantitative analysis. The study was approved by our institution's Institutional Review Board, and all participants provided informed consent before participating.

\subsection{Conditions}

Our study had three conditions provided by the same mixed reality headset:

\paragraph{\textbf{Conversation practice videos (control)}} Participants watched expert-curated conversation practice videos. To make the experience more similar to a roleplay, participants were instructed to shadow one speaker's lines by repeating after the speaker and to pause the video after questions and produce their own answer first. Participants were welcome to repeat and review any parts of the video.
 
\paragraph{\textbf{\systemname preset (treatment)}} The study participants used \systemname with a predetermined set of learning objectives for each context, derived from the conversation practice videos described above. Although the inputs were always the same for each context, \systemname generated a new set of scenes, characters, and dialog every time it was used. Participants were asked to complete the roleplay of all scenes generated by \systemname.
 
\paragraph{\textbf{\systemname create your own (CYO)}} Participants created custom scenarios, specifying their own setting, tasks, grammar, and vocabulary or having \systemname generate them. They practiced their custom scenario with the option to stop early at any point.

\subsection{Procedure}

Because each participant underwent both controlled conditions, we prepared two contexts: Cafe and Shopping. Participants were assigned to one of four counterbalanced groups determining which condition they experienced first and which context was used for each condition. Both conditions were delivered with the same Apple Vision Pro headset.
 
Each participant completed the following sequence: pre-test and pre-survey for the first context, a headset tutorial, the first intervention (control or treatment depending on group), post-test and post-survey, an optional break, then pre-test and pre-survey for the second context, the second intervention, and post-test and post-survey. After both controlled conditions, participants created and practiced their custom scenario (CYO), completed a post-survey, and participated in a semi structured interview. The sessions lasted two to three hours. The participants were compensated with a \$75 USD gift card.

\subsection{Measures}

We measured learning outcomes, confidence, narrative engagement, presence, sense of agency, and support quality using the following instruments.

\paragraph{\textbf{Knowledge test.}} Participants completed pre- and post-intervention knowledge tests with two versions (Cafe and Shopping contexts), each consisting of a vocabulary quiz and a speaking test. The vocabulary quiz asked participants to recall five Japanese words given English meanings (scored 0-2 per item, range [0,10]). The speaking test used structured roleplay in which participants heard Japanese dialog, read English response instructions, and responded aloud in Japanese until completion.
A native Japanese-speaking researcher with second language acquisition experience wrote the roleplay scripts using the learning goals for each context, avoiding exposure to either condition to prevent bias. The speaking test assessed task completion and language use for four goals (scored 0-2 per goal, range [0,8]). Both tests were graded by two Japanese-speaking researchers blind to participant identity, condition, and timing, with excellent inter-rater reliability: ICC(2,k) = 0.990 for vocabulary and ICC(2,k) = 0.971 for speaking.
 
\paragraph{\textbf{Confidence.}} Immediately after each knowledge test, participants rated on five-point Likert scales how confident they felt about their ability to converse in Japanese in the corresponding context and whether the tool used in that condition could help them prepare.
 
\paragraph{\textbf{Narrative engagement.}} We adapted 11 items from the Narrative Engagement Scale~\cite{busselle2009measuring} to assess four subscales: narrative understanding, attentional focus, narrative presence, and emotional engagement. Each item was rated on a five point Likert scale. \footnote{We removed ``My understanding of the characters is unclear'' to focus our measurement of narrative understanding on the story, not the characters.}

\paragraph{\textbf{Presence.}} We included items on social presence and spatial presence from the New Presence Scale ~\cite{vhil_new_presence_scale_2022}. Each item was rated on a five point Likert scale.
 
\paragraph{\textbf{Support evaluation.}} Two Japanese-speaking researchers coded all real-time corrective feedback and post-reflection lesson feedback from system logs, assessing whether feedback was warranted, whether it correctly identified errors, whether it modeled the correct answer, and whether metalinguistic explanations were appropriate.

\paragraph{\textbf{Log coding.}} Two researchers coded the session logs of all 43 participants (232 scenes, 841 assigned goals). For each goal assigned to a scene, the authors coded whether the scene created an opportunity to practice it and whether the learner made a recognizable attempt, recording reasons for missed opportunities. For each of the 148 scene transitions within a session, they coded whether the new scene meaningfully continued the previous one, was compatible but stand-alone, broke continuity, or was unclear, recording break types where applicable. Disagreements were reconciled to consensus.
 
\paragraph{\textbf{Interview.}} A semi-structured interview at the end of the study asked about the general experience of the participants, comparisons between conditions, and impressions of \systemname's coherence, support, and potential for regular use. The interviews were transcribed and analyzed using thematic analysis~\cite{BraunClarke2006}. Two researchers developed an initial codebook, calibrated in five interviews, and then divided the remainder.

\section{Results}
Below, we present the findings from our study, organized by their relevance to our 3 research questions.

\subsection{Learning and Support (RQ1)}
\label{sec:findings:learning}

Below, we examine learning outcomes relative to expert-authored practice targeting the same goals, whether generated scenes provided opportunities to practice those goals, and the quality and use of the generated support.

\subsubsection{\textbf{Vocabulary recall and confidence were higher after GELLI; speaking performance did not differ significantly.}}
\label{sec:findings:learning-outcomes}

In the controlled comparison ($N = 40$), participants improved from pre- to post-test in both conditions. In the vocabulary quiz, which assessed the recall of five target words per context, the scores increased in both the video condition ($M = 4.40$ to $7.75$, $t(39) = 7.95$, $p < .001$, $d = 1.21$) and the GELLI condition ($M = 4.40$ to $8.60$, $t(39) = 10.79$, $p < .001$, $d = 1.78$). A linear mixed-effects model that predicts the post-test score from condition, context, and pre-test score, with a random intercept for the participant, found significantly higher recall after GELLI ($b = 0.85$, $SE = 0.31$, $z = 2.79$, $p = .005$, 95\% CI [0.25, 1.45], $d = 0.45$). Confidence in conversing in the practiced context also increased in both conditions (video $d = 0.78$; GELLI $d = 1.02$) and was significantly higher after GELLI ($b = 0.59$, $SE = 0.19$, $z = 3.18$, $p = .001$, 95\% CI [0.23, 0.95], $d = 0.45$).

The speaking test assessed task completion and use of target vocabulary and grammar in a structured roleplay. Task-completion scores improved in both conditions (video $d = 0.52$; GELLI $d = 0.58$), with no significant difference between-conditions ($b = 0.15$, $SE = 0.19$, $z = 0.79$, $p = .43$). Vocabulary and grammar use also improved for both conditions (video $d = 0.99$; GELLI $d = 1.03$), again with no significant difference ($b = 0.46$, $SE = 0.41$, $z = 1.13$, $p = .26$).

\subsubsection{\textbf{Generated scenes created opportunities for most assigned goals, and learners took most of them.}}
\label{sec:findings:goal-opportunities}

Of 841 goals assigned in 232 scenes, 760 (90.4\%) had provided an opportunity to practice in the scene, and the learners attempted 622 (74.0\%) or 81.8\% of the goals for which they had an opportunity. Communicative tasks were both more reliably staged (95.3\% with an opportunity) and most often attempted (89.6\%). Vocabulary goals were attempted at 70.4\% and grammar goals at 51.4\%. The 81 goals without an opportunity fell into two main patterns, which often co-occurred: the character pre-empted the act the goal required (53.1\%), for example by volunteering information the learner was meant to ask for, or the conversation offered no clear opening for it (51.9\%). Scenes ending before the goal could arise (8.6\%) and goals too broad to stage (3.7\%) also occurred. 

\subsubsection{\textbf{Support windows let learners explore and discover language, and attempt more than they would by default.}}
\label{sec:findings:support-experimentation}

\paragraph{Support provided security} Learners described feeling anxious or uncertain during conversations, but that having support available helped them continue. P012 explained, ``having the transcripts and suggestions gave me something to lean on, so I didn't just freeze. It made me feel more reassured and confident.'' The way feedback was delivered also mattered. P044 said ``It also reduced stress around mistakes, since it corrected me gently [...] like saying that the girl was confused.'' P036 described this behavior as making them feel like ``the system was helping [them] figure it out.'' 

\paragraph{Suggestions gave learners options to explore how to express themselves.} The suggestions provided building blocks while leaving the learners to construct their own responses, consistent with the design choice to offer phrases rather than full responses (DG5). P020 shared that ``this type of learning, I think it kind of encourages you to come up with the whole sentences yourself, and then it provides me with some of the useful phrases that I might need.'' Having several options helped learners consider different ways to express themselves. P029 valued the notes because ``It gives you so many different options to choose from.'' P017 connected the suggested vocabulary to wanting to ``express myself with the exact words I wanted,'' describing the experience as ``a lot of freedom.''

\paragraph{Suggestions prompted discovery and experimentation with new language.} Many participants valued how suggestions surfaced unfamiliar language they could try on the spot. P026 described ``getting exposure to like new vocabulary along the way, as my conversation naturally evolved'' as the most enjoyable part of the experience. Encountering unfamiliar suggestions could initially be difficult but also prompted experimentation. P021 described this in detail: ``it like forced me to be more proactive about using new vocabulary, especially because sometimes there would be completely unfamiliar things in there. So, at first it was frustrating and confusing to see it, but once I actually said it like in practice or listened to someone else saying it, it was pretty helpful.'' P035 described the same friction: ``Sometimes it responded with vocabulary I didn't know, like I just wanted basic words, but it used advanced terms. That was frustrating, but also kind of fun.'' Opportunities for reflection extended this experimentation beyond the immediate response, allowing learners to revisit what they had said and consider alternatives. P038 explained, ``This tool lets you experiment with different phrases and ask questions at the end, like `Could I have said it this way instead?'\,''

\paragraph{Support helped learners attempt more than they would by default.} In all of these accounts, the effect of support was that learners tried things they would otherwise have avoided. P046 explained, ``without the support, I definitely would not have picked answers that were as ambitious as I did.'' and described the support as there ``so I can use it if I want, or similarly to learn a little more than what I would just default to.'' P005 compared the experience to their time in Japan: ``I often didn't understand everything and gave short responses. But here, I responded more because I had time to think, and the system offered me options for what to say.'' P021, reflecting on the unfamiliar vocabulary in the suggestions, concluded: ``I feel like I was pushed more outside of my comfort zone to learn more.''

\subsubsection{\textbf{Generated feedback was mostly warranted and reliably modeled corrections.}} We evaluated whether generated feedback addressed actual errors and provided useful explanations, and whether post-scene lessons were relevant and appropriate. Of 264 generated feedback instances, we excluded 110 involving possible pronunciation or speech-recognition errors. Because audio recordings were not retained, we could not distinguish learner errors from transcription errors in these cases. Of the remaining 154 instances, 137 (89.0\%) responded to actual errors and 17 (11.0\%) were false positives. Among the 137 warranted instances, 99.3\% modeled the correct answer, 97.1\% provided an appropriate example, and 97.1\% included a metalinguistic explanation. Of those explanations, 59.4\% supplied sufficient rationale and 40.6\% were cursory. Of 228 evaluated post-scene lessons, excluding eight with missing data, 78.1\% were relevant to the participant's reflection or conversation, 19.3\% somewhat relevant, and 2.6\% irrelevant; 72.4\% were appropriate, 27.2\% somewhat appropriate, and 0.4\% inappropriate.

\subsubsection{\textbf{The amount of support could initially feel overwhelming.}}
\label{sec:findings:support-overload}

The amount of available language could make it difficult to decide what to use. P026 described ``choice paralysis'' from the vocabulary appearing during interaction. P031's account suggests that initial overwhelm could ease as they became familiar with the interface. They initially scrolled through tabs without retaining anything, but ``once I realized I could build up gradually, in the second and third scenarios I started picking vocab and phrases, combining them, and making progress.'' They described the information becoming ``useful, instead of just noise.''

\subsection{Immersion in the Generated Narrative World (RQ2)}
\label{sec:findings:immersion}

Below, we compare narrative engagement and presence across conditions, then examine how representations, character responses, and learners’ imagination supported immersion and what disrupted it.

\subsubsection{\textbf{GELLI produced greater presence and emotional engagement, but lower narrative understanding.}}
\label{sec:findings:immersion-measures}

We compared narrative engagement and presence using Wilcoxon signed-rank tests, Holm--Bonferroni corrected across six subscales. Participants reported significantly greater narrative presence ($W = 73.5$), emotional engagement ($W = 53.0$), social presence ($W = 31.0$), and spatial presence ($W = 52.0$) with GELLI than with the videos (all corrected $p < .001$). Attentional focus did not differ significantly ($W = 219.0$, corrected $p = .57$). Narrative understanding was significantly higher in the video condition ($W = 28.5$, corrected $p < .001$). To examine whether the immersion and engagement learners reported were associated with learning, we fit linear mixed-effects models predicting vocabulary post-test scores from each presence and engagement subscale, controlling for condition, context, and pre-test score, with a random intercept for participant. None of the four predictors (narrative presence, spatial presence, social presence, emotional engagement) significantly predicted post-test scores. 

\subsubsection{\textbf{Static imagery, narration, and responsive characters produced a sense of place and of someone there.}}
\label{sec:findings:immersion-features}

\paragraph{The narrative representation established where the interaction was and who was there.} As P012 explained, ``The images set the scene. Seeing a café background or a character up close helped with immersion. Layering was especially effective; having a background plus a person in front made it feel like someone was really there. The character in front was the most helpful for making it feel real.'' P022 described how the images and narration worked together: ``they described the tables were, like, next to the windows, and I could, like, literally see that. [...] so you might not understand everything the person is saying, but the fact that you can see it helps the context.'' Some learners attributed immersion to the way the headset presented the scene and allowed them to participate. P037 explained that ``with the headset on and my hands free, it felt like I was really conversing in the moment.'' P040 emphasized isolation from the surrounding environment: ``Looking in a VR headset versus a computer screen blocks out distractions; the size and enclosure of the images help immersion.''

\paragraph{Responsive, believable characters deepened immersion.} Characters felt real when their behavior fit their roles and responded to what learners contributed. P037 explained, ``even if I didn't give much detail, it understood what I meant, which made it feel organic, like talking to a native who can follow context. And when it couldn't figure out what I meant, it asked for clarification, just like a native speaker would.'' P004 felt included when a character recalled their name: ``I was like, whoa. It knows who I am [...] it feels like you're included, like, included in this world. It feels like you get to, like, shape, the direction of where it goes.'' The ability to depart from the assigned task could also make the conversation feel real. P017 ``felt like the characters were real [...] because the content was constantly being generated based on what I responded,'' especially when they went ``off track'' and complimented a character's hand: ``it felt like they were real because, like, they responded to some off track answer.'' 

\paragraph{Social presence created pressure to respond.} Being addressed by a character produced a familiar pressure that learners described as productive. P038 explained, ``it felt like the person in front of me was expecting me to respond. You feel pushed to try your best'' P035 said, ``It felt more special, like I owed something to someone [...] when the character responded, I felt compelled to continue.'' P003 described it as ``less intimidating compared to talking to a real person'' but ``still kind of a little bit stressful, which is good, because it mimics the real like scenario.''

\subsubsection{\textbf{Immersion involved learners' imagination, supported by images, narration, and their own experience.}}
\label{sec:findings:imagination-narration}

\paragraph{Learners described immersion as imagination.} The study participants described using their imagination to bring static images to life. P038 said, ``I found myself imagining facial expressions or reactions, so [static images weren't] really a problem for me,'' and P024 noted, ``even if it was just like a still image, for some reason, I was imagining like a lot like, it's talking and stuff.'' P040 said, ``You have to rely on imagination. The images and VR headset put me in the middle of it, which helps me engage without being overly aware that I'm pretending.'' Narration contributed to this: P008 valued descriptions of a character ``nodding or like taking notes'' because they helped them ``imagine what's happening in like a world that it's created.'' P016 initially found the narration excessive, ``but I feel like when I am in the conversation, I understand why the narrative is there. Think it really creates like, Oh, you're actually in that environment talking to this real person.'' Narrated reactions could also prompt a response. P018 said, ``when the narrator says, like, oh, this person slightly tilted her head or like, when they say like, oh, sorry. Can you repeat? That makes me personally feel like, Oh, I'm so sorry. Let me say it again.''

\paragraph{Imagination could often require a conscious effort.} P037 put it as a choice: ``It's hard to perfectly replicate a real-world feeling. But if you immerse yourself voluntarily, I think you can get a lot out of it.'' P012 similarly said, ``You have to believe you're in the scenario. I reminded myself I wasn't actually talking to someone so I could relax enough to try.'' P043 described making that decision and what followed from it: ``once I accepted that the purpose was just to simulate real-world practice, I relaxed and treated it like a real conversation. As I got used to it, I actively thought, `If I were really in a coffee shop right now, what would I say?''' 

\paragraph{Many imagined situations they had been in.} Their memories provided learners with material for imagining the encounter. P037 ``used my own memories of being in those scenarios, which made it immersive,'' while the visible character gave them ``a person to talk to.'' P006 said, ``I've lived in Japan for two years, and it felt very real—it reminded me of my experiences there.'' P023 recalled restaurant, café, and shopping interactions from studying abroad, including being asked ``literally the same questions,'' and suggested drawing on analogous experiences: ``if you don't have experiences like that, then you can also translate ones that you already have that occur in different languages and contexts to it too.'' 

\subsubsection{\textbf{Artifacts of generation interrupted immersion.}}
\label{sec:findings:immersion-disruptions}
P037 described changes in a character's appearance across scenes: ``Sometimes the style was very anime-like, then suddenly more realistic, even when it was supposed to be the same person. [...] That breaks immersion a bit; it feels like, `Oh right, I'm in an AI world.'\,'' P008 felt that three scenes set in the same classroom appeared to take place in ``three different classrooms'' because the backgrounds differed. Images could also suggest unintended roles: P007 interpreted a caf\'e character seated across the table as a date rather than a waitress. P044 found sudden changes in vocal pitch distracting and wanted voices that better matched the character and role.

\subsection{Narrative Co-Authoring (RQ3)}
\label{sec:findings:coauthoring}

Below, we examine how learners authored scenarios and shaped conversations, and how generated encounters held together across scenes.

\subsubsection{\textbf{Learners valued authoring scenarios for both preparation and play.}}
\label{sec:findings:authoring-preparation-play}
\begin{figure}[t]
    \centering
    \includegraphics[width=\linewidth]{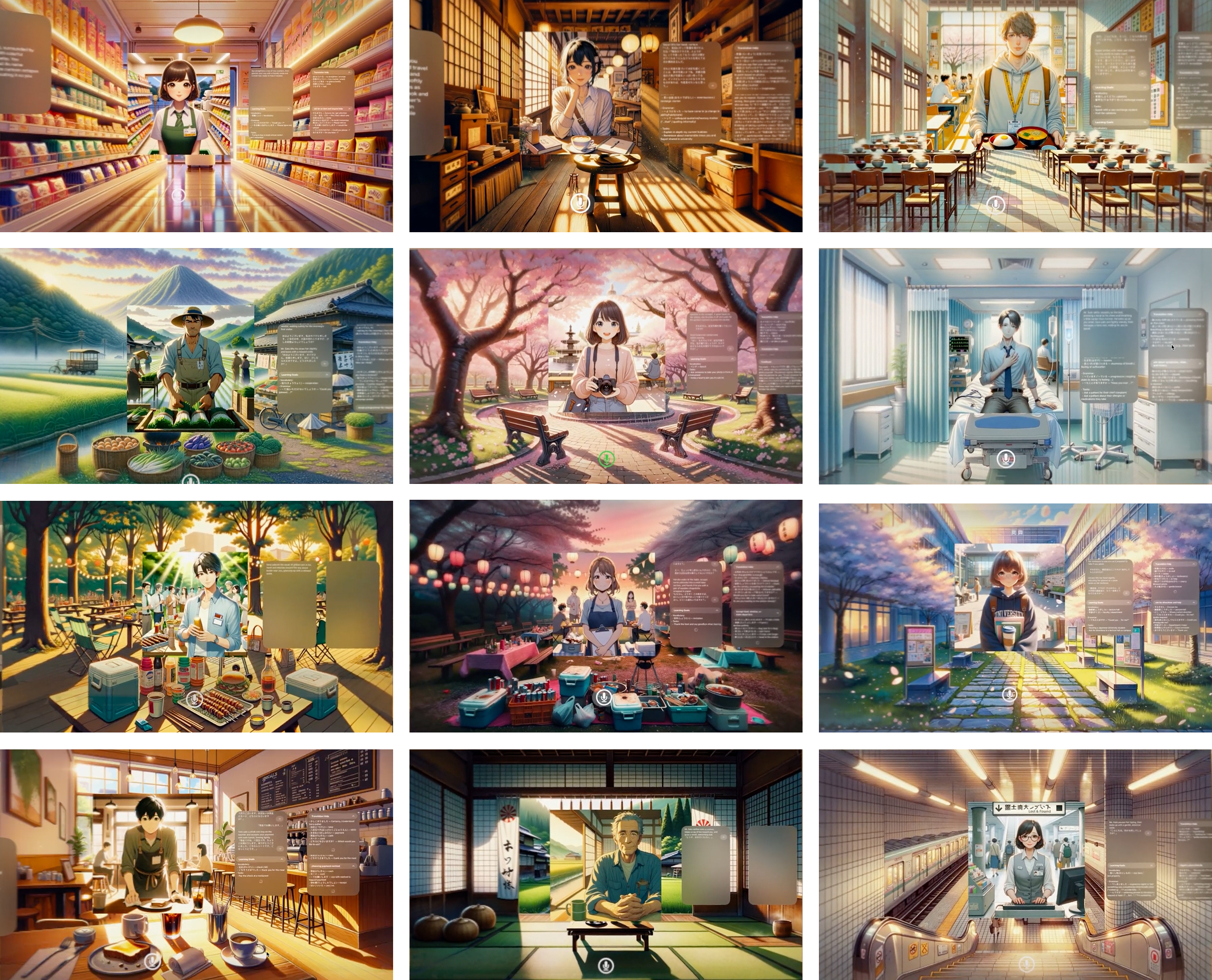}
    \caption{Examples of scenes generated from participants' custom scenarios, illustrating the range of situations they chose for language practice.}
    \Description{A grid of twelve screenshots showing generated characters and environments, including shops, a classroom, a park, a clinic, a barbecue, and a train station.}
    \label{fig:cyo}
\end{figure}

Figure~\ref{fig:cyo} illustrates the variety of scenes generated from participants' custom scenarios.

\paragraph{Scenarios felt useful for preparing for real life.} Participants' confidence in conversing in their self-chosen settings increased from before to after the custom scenario ($M = 2.44$ to $3.15$, $p < .001$, $d = 0.72$). They also reported high confidence that GELLI could help them prepare for those settings ($M = 4.23/5$). Participant reports corroborate these results, as learners described wanting to use GELLI to prepare for future real-life encounters. P041 said ``I could model any situation. Relive past conversations or try new ones. That was the most fun.'' P022 envisioned using GELLI to emulate specific situations and receive feedback on ``what sounds natural and what people would say in those situations.'' P033 emphasized targeting particular learning needs ``You can tailor lessons to your weak points in vocab or grammar, and the AI generates fresh scenarios.'' P016 said the scenario they created ``is exactly like what I would envision myself use Japanese for.''

\paragraph{Participants thought GELLI could be used as both play and practice.} Beyond preparing for real life, participants envisioned practicing through fantastical or impossible situations. P040 explained, ``Putting myself in a fictional world motivates me and keeps me interested in learning.'' Other learners imagined playful scenarios they would like to try. P036 proposed an amusement park or dinosaurs as occasions to learn new vocabulary. P031 was now thinking about ``going to the moon and meeting a Japanese person there.'' For some, their interest in experimenting with a generated encounter was part of what made them want to practice. P022 explained, ``I want to explore, like, what this tool can do, and [...] practice the Japanese, like language while I'm at it.'' They described exploring the different ways the tool could be used both for learning and for ``having fun with it kind of like a game.''

\paragraph{Learners could struggle with specifying appropriate scenarios} P012 attributed a break in their café scenario to the prompt: ``I only wrote `café' in the prompt [...] So it's important to write very specific prompts.'' P046 unintentionally ``picked something that was harder [...] something very much outside my comfort zone.'' Two learners wanted authoring tied to prior study: P012 said that ``Without that prep, it's harder to know what to focus on in the moment,'' and P035 that ``it would be less overwhelming if it tied into a class.''

\Needspace{4\baselineskip}
\subsubsection{\textbf{Learners felt they were co-authoring the unfolding encounter with GELLI.}}
\label{sec:findings:shared-authorship}

\paragraph{Learners could steer, and the narrative followed.} Learners described shaping the narrative together with the model. P044 described their fan meeting scenario as ``open-ended and felt like we were creating the story together.'' P009 described the dialogue as ``shaped by both of you, like both the model and you.'' Learners also tested this system deliberately. P041 said, ``I was purposely changing the subject, and it handled it well,'' and P025 started ``shopping for, like a shirt, and then adjusted to shoes or something like that. And it was completely, like, perfectly adjusting to the conversation.'' P010 enjoyed how GELLI would ``roll with me and kind of customize the scenario,'' and P002 noted feeling like the model was ``guiding you through it, which I do think is helpful.''

\paragraph{The system's steering toward goals could sometimes interfere or feel disruptive.} The same guidance could be disruptive and remove reasons to speak. P039 described a custom encounter as coherent and aligned with their goals while finding it less natural: ``I could tell the story was structured around the learning goals we set.'' P045 said, ``it kept steering toward the main goal, like signing the waiver or paying the fee. It worked, but it felt less conversational [...] Still, I get that you need stopping points.'' P020 noted that the character sometimes supplied information they had intended to request. In a gym-membership encounter, it would ``slip in information that I was about to ask for in my next question,'' which was frustrating ``because I do want to practice that.'' P010 described the character's in-role confusion arriving ``when I was doing actions that didn't align with the learning goals,'' so that ``the learning goals were the direction it was trying to go with the conversation, which may or may not be the same as where I was trying to go.''

\subsubsection{\textbf{Generated narratives held together across scenes, with few breaks.}}
\label{sec:findings:continuity}

\paragraph{Scenes unfolded into believable sequences, sometimes in directions learners had not expected.} Participants described coherence through sequences of events that unfolded in ways they felt were realistic, even when they could be unexpected. In P046's account, losing a wallet also meant losing the transit card inside it, creating a reason to buy a ticket in a later encounter. They had not expected the story to begin at lost and found, but appreciated how the system connected that event to the rest of the experience. P022 recreated a barbecue with a Japanese community to practice discussing their work with new acquaintances. GELLI developed this starting point by generating several characters with whom they had unique conversations. The narrative eventually ended with a scene in which everyone left while they offered to help clean up, which surprised them: ``I was very surprised that it was able to come up with that kind of closing scene, like a closing scene where people are starting to leave. I was very like, impressed that the last scene, while it wasn't necessarily about the prompt, it still closed out the scene that the prompt entailed. So, yeah, it did feel like it was very cohesive'' They noted that this progression was helpful because they could ``learn all of those interactions in one go.'' P031 also commented on the generated elements expanding on their prompt: ``The AI improved on it—it assumed I'd ask a train conductor instead of a random person, which made more sense. At the Sky Tree, I said I'd ask where to buy tickets, and it put me at the ticket counter talking to a staff member. That was smarter than what I thought of.'' Within the scenes, continuity also held. P036 said, ``She stayed in the same spot, remembered what I'd said, and kept track of context. When I said, `I'll take the pasta,' she gave it to me, and then reminded me that the natto I was looking for was still back there for the recipe.'' 

\paragraph{Breaks in continuity were rare, and mostly repetitions.} Across the 148 scene transitions within sessions, 105 (70.9\%) meaningfully continued the previous scene, 31 (20.9\%) were compatible with it but stand-alone, four (2.7\%) broke continuity, and eight (5.4\%) were unclear. All four breaks involved repeating an act that the learner had already completed, and one also shifted the premise or the learner's role. Learners' accounts matched these patterns. P019 reported being asked to invite a character on a date again after already doing so, and P023 described starting as someone visiting another person, but later being the holder of an appointment and eventually receiving an X-ray. Stand-alone scenes fit the setting without building on prior events, which is what P012 recognized when their custom café scenario ``suddenly shifted to ordering café items instead of continuing my conversation with Akira. That broke the flow.'' Learners tended to tolerate smaller lapses: P011 noted ``he forgot my name'' but found that the narrative ``overall, is pretty continuous.''

\section{Discussion}
\label{sec:discussion}

Our findings suggest that \systemname's language practice generated at runtime can support learning and narrative immersion while giving learners a meaningful role in authoring and steering their experience. We discuss these findings in relation to learning goals and narrative agency, immersion in generated worlds, scaffolding, and learner authoring, then draw design implications for generated practice.

\subsection{Learning Goals, Narrative Coherence, and Learner Agency}
\label{sec:discussion:alignment}

Interactive narrative learning environments must manage learner agency while guiding them toward learning goals and keeping the narrative coherent \cite{oakstory, riedl2013interactive}. Our findings show how this coordination can work at runtime and where it can break down.

\subsubsection{Providing targeted practice while preserving learner agency.}
Expert-authored practice can guarantee that the activities learners complete are aligned with their learning goals, whereas high-agency learning environments risk leaving learners without enough structure to make progress toward them \cite{oakstory}. Open-ended LLM dialogue can drift from intended tasks or fail to elicit target language forms \cite{ellmaT, glandorf2025grammar}. Our coding suggests that focused narrative scenes with adaptive support provided opportunities for most goals without constraining what learners said (Section~\ref{sec:findings:goal-opportunities}). Uptake was highest for communicative tasks, lower for vocabulary, and lowest for grammar, suggesting that scenes could prompt communicative acts without necessarily eliciting the specific language targeted for practice. Learners' exploration of unfamiliar forms encountered in support, or alternative directions for the narrative, may have contributed to this pattern (Sections~\ref{sec:findings:support-experimentation} and \ref{sec:findings:shared-authorship}).

Learners nevertheless recalled more target vocabulary and reported greater confidence than after expert-authored practice, while speaking performance improved in both conditions with no significant between-condition difference (Section~\ref{sec:findings:learning-outcomes}). Exposure to target words in character dialogue and support windows may have contributed to the vocabulary advantage even when learners did not produce those words; our coding did not assess this possibility, which warrants further study. Together, our findings suggest that targeted practice need not come at the cost of agency and interactivity.

\subsubsection{Keeping generated narrative coherent while steering toward learning goals.}
Prior work on experience management addresses the tension between a coherent narrative and a participant's ability to change its direction \cite{riedl2013interactive}. In a learning environment, the system must also steer that narrative toward learning goals. Our approach of generating successive scenes around preceding events and remaining goals often produced connected encounters, including developments learners had not anticipated but found believable (Section~\ref{sec:findings:continuity}).

Within scenes, however, a helpful, believable character response could advance the encounter while removing the learner's reason to perform the intended communicative act, for example by volunteering information they had intended to request. Conversely, steering the conversation toward unfinished goals could interfere with the learner's direction and make the exchange feel less natural (Section~\ref{sec:findings:shared-authorship}). Preserving practice opportunities therefore requires coordinating what makes sense for the narrative with what gives the learner a reason to use the language. Resolving this tension may require better predictions of what the learner intends and how the narrative might unfold, and room to move goals to later scenes rather than forcing them into the current one.

\subsection{Immersion in Generated Narrative Worlds}
\label{sec:discussion:immersion}

Bruni et al. distinguish system immersion, produced by the technology, from narrative immersion, produced by imagery, characters, and storytelling, and argue that the second can be evoked largely independently of environmental fidelity \cite{bruni2022digital}. Our findings suggest that learners experienced narrative immersion without an authored plot or an explorable environment.

\subsubsection{Narrative immersion with lightweight representations.}
Presence research finds that perceived realism depends on whether an environment meets the user's expectations, not only on how it looks \cite{slater2009place}. Learners reported greater narrative presence with GELLI than with videos on the same headset (Section~\ref{sec:findings:immersion-measures}). Their accounts linked immersion to agreement among imagery, narration, character responses, and expectations, with inconsistencies disrupting the experience (Sections~\ref{sec:findings:immersion-features} and \ref{sec:findings:immersion-disruptions}). These findings suggest that lightweight representations can support immersion when what learners see, hear, and do fits together, making consistency across generated scenes a design priority.

\subsubsection{Narrated non-verbal cues.}
Embodied conversational agents convey understanding and coordinate conversation through gaze, gesture, and expression \cite{gelliCassell2001}. Studies of LLM-based embodied agents for language learning report that these cues matter to learners \cite{ellmaT, ellmaTchi}. In our study, learners used narrated gestures and reactions to picture character behavior and, in some cases, decide how to respond (Section~\ref{sec:findings:imagination-narration}). Our findings suggest that narration can make social cues available for interpretation and response without requiring each behavior to be animated.

\subsection{Scaffolding Without Taking Over}
\label{sec:discussion:support}

The involvement-load hypothesis emphasizes searching for word meanings and evaluating words in context in vocabulary learning \cite{laufer2001incidental}. Research on computational feedback identifies a risk of relying on feedback instead of evaluating one's own work \cite{candc2024feedback}, while studies of LLM tutors report difficulties with long, generic, or excessive corrective feedback \cite{ellmaT, ellmaTchi}. Our findings show why support must balance involvement with security. Having support available helped learners feel confident enough to attempt unfamiliar language and more ambitious responses. Offering phrases rather than full responses left them responsible for constructing what they said, using suggestions to express their intentions and explore unfamiliar language and narrative directions (Section~\ref{sec:findings:support-experimentation}). Yet some learners found the options tiring or difficult to choose among (Section~\ref{sec:findings:support-overload}). This calls for support that keeps exploratory options available while allowing learners to control how much they see.

\subsection{Authoring in Emergent Narrative}
\label{sec:discussion:authoring}

Emergent narrative shifts authoring from writing a plot to defining the characters and circumstances from which a story can develop through interaction \cite{aylett1999aaaiFS, louchart2004ijceell}. Generative learning systems have begun to give learners a hand in this authoring \cite{peng2023storyfier, ellmaT}. Learners authored scenarios for preparation and play, but sometimes struggled to specify the situations they imagined.

\subsubsection{Learners as authors of practice situations.}
In our study, learners often created situations they expected to face, such as a barbecue with a Japanese community, and situations they wanted to play in, such as a fantasy world. Confidence in conversing in their self-chosen situation rose after practice, and learners rated \systemname as likely to help them prepare for it (Section~\ref{sec:findings:authoring-preparation-play}). These findings concern perceived preparation; transfer to real conversations remains untested. Learners also described the encounters as something they and the system created together (Section~\ref{sec:findings:shared-authorship}).

Research links imaginative suggestibility with presence in VR \cite{jicol2023imagine}. Our findings suggest that authoring may support imaginative participation by letting learners define situations from remembered or anticipated encounters and bring expectations against which they interpret the generated world (Sections~\ref{sec:findings:authoring-preparation-play} and \ref{sec:findings:imagination-narration}). Their accounts of immersion when characters and events fit their expectations suggest that authoring can help establish a situation they find believable. Yet learners could struggle to convey their intended situation, as some specified too little and others created scenarios that were too difficult for them to understand (Section~\ref{sec:findings:authoring-preparation-play}). Authoring support should therefore help learners see and revise how the system interprets their setup, so that the generated situation is both suitable and remains recognizable as it develops.

\subsubsection{Practice as experimentation.}
Experiential learning theory links experience, reflection, abstraction, and active experimentation \cite{Kolb1984}; Morris emphasizes contextually rich, unfamiliar experiences in which learners make decisions and examine their understanding \cite{Morris2020ILE}. Taken together, our findings suggest that choosing a situation, imagining themselves within it, and having support to try unfamiliar responses made practice a way to explore possibilities, test expectations, and discover further things they wanted to learn. Generation extended these possibilities by introducing language and developments learners had not specified and responding to their choices. Authoring gave learners responsibility for defining their experience, while responding to generated events required them to adapt that initial intention. Learners could therefore experiment with what to say, which situations to enter, and how the narrative would unfold (Sections~\ref{sec:discussion:support}, \ref{sec:discussion:authoring}, and \ref{sec:discussion:alignment}). Letting learners revisit an encounter would extend this experimentation by making alternative choices and their consequences available for comparison. Reflection and abstraction remain less clear: we could not determine how deeply learners engaged with them (Section~\ref{sec:limitations}).

\subsection{Design Implications for Generated Practice}
\label{sec:discussion:implications}

We draw seven design implications from the findings above, grouped by the problem they address. Each names the sections that motivate it.

\subsubsection{\textbf{Steering practice toward learning goals.}}
\leavevmode\par\noindent

\textbf{Design Implication 1: Model how narrative developments create and remove practice opportunities} (Section~\ref{sec:discussion:alignment}). Systems should combine a model of the unfolding narrative with a learner model that tracks what the learner appears to know, what they have practiced, what remains, and what they appear to expect next. Planning should consider several exchanges ahead and make revisions as learners act. Where a goal fits the anticipated direction, the system can place an opportunity along that path; where it does not, it can introduce a plausible development that gives the learner a reason to pursue it.

\textbf{Design Implication 2: Let goals move beyond the scenes initially assigned to them} (Section~\ref{sec:discussion:alignment}). If an exchange removes the learner's reason to attempt a goal, the system should retain the goal and establish an opportunity elsewhere in the narrative. Intermediate exchanges can create new reasons to use the language, and additional scenes or sessions provide space for them. When exploration reveals a new learning interest, the system could suggest an additional goal while retaining unfinished ones.
    
\Needspace{4\baselineskip}
\subsubsection{\textbf{Representing generated narrative worlds.}}
\leavevmode\par\noindent

\textbf{Design Implication 3: Design for consistency between representation, narration, and interaction before adding fidelity} (Section~\ref{sec:discussion:immersion}). Generated imagery, narration, and character behavior should agree with each other and with the situation the learner specified. Narrated details that are visible should match the image, and details the image cannot show, such as a character's actions, can be carried by narration alone. Characters should keep the same appearance across scenes, and their responses should fit their role and what the learner has said. Where fidelity and flexibility conflict, favor a representation that generation can keep consistent as scenes change.

\textbf{Design Implication 4: Use narration to make unseen actions available for participation} (Section~\ref{sec:discussion:immersion}). When deciding how to represent a character's action, consider what the action communicates and how the learner can respond to it. Narration can describe hesitation, acknowledgment, or activity beyond the visible scene, giving learners something to interpret and act on. Connect those descriptions to the character's subsequent behavior so that learners' responses have consequences.

\subsubsection{\textbf{Supporting authoring and exploration.}}
\leavevmode\par\noindent

\textbf{Design Implication 5: Offer support as a way to think through how the encounter could unfold} (Sections~\ref{sec:discussion:support} and \ref{sec:discussion:authoring}). Use the conversation so far and predictions about where it could go to offer a small set of options that represent different ways of responding and the different directions each would take the encounter. Let learners adjust how many options they see and hide or expand them as needed, while preserving their responsibility for selecting, adapting, and combining them into a response. Provide lookup and explanations for intentions the system has not anticipated.

\textbf{Design Implication 6: Let learners revisit decisions and compare alternative continuations} (Sections~\ref{sec:discussion:support} and \ref{sec:discussion:authoring}). Allow learners to return to an earlier exchange, try a different formulation or action, and see how the conversation develops. Connect this to reflection on why language worked in that context and how another intention would call for a different response. This lets learners examine possibilities they could not pursue in the same exchange and identify further expressions or situations to practice.

\textbf{Design Implication 7: Make the generated interpretation of the learner's scenario visible and revisable} (Section~\ref{sec:discussion:authoring}). Build on scenario previews by helping learners specify and revise the roles, relationships, circumstances, and intended interactions that matter to them, since a setting alone leaves these ambiguous. Allow learners to correct the system's interpretation before and during participation, and carry those corrections into subsequent dialogue, imagery, and narration. Authoring support can also identify goals that need additional narrative setup and suggest how they might fit the learner's scenario.

\section{Limitations and Future Work}
\label{sec:limitations}

\subsection{Attribution}
Our design bundles interactivity, generated imagery, narration, voice, support, and reflection into a single treatment, and the expert-authored videos were chosen as a reference for pedagogical alignment rather than for interactivity. The comparison found higher vocabulary recall and confidence after generated practice, with no significant difference in speaking performance. It does not establish which components produced these outcomes or whether generation would outperform an interactive authored alternative. The absence of a significant association between presence and vocabulary recall does not establish the relative contributions of representation, support, and reflection; isolating these contributions would require further comparisons. We also designed support to carry learners through reflection and abstraction, but reflections had no required content and we did not measure how closely learners read the generated lessons, so we cannot confirm that these phases occurred as intended. Our log coding was reconciled to consensus between two researchers rather than independently rated, so we do not report inter-rater reliability.

\subsection{Duration and Transfer}
The controlled comparison used two contexts in a single session. We cannot claim that the learning advantages persist, transfer to real conversations, or generalize to other contexts. The presence and engagement learners reported may matter more for sustained practice than for immediate gains, but testing this requires a delayed post-test or a longitudinal study. A natural extension would track learners over multiple sessions with a user model that reincorporates missed goals, generates new situations as proficiency develops, and spaces review based on what has been practiced. Learners' reports of preparation and transfer were self-reported, and future work should test performance in real conversations.

\subsection{Scope and Generalization}
Our participants were adults in one area learning Japanese, and results may differ for other populations, proficiency levels, or languages. The architecture is largely language agnostic, but Japanese required specific handling of its writing system, furigana, and register, and low-resource languages would need additional work. We did not test a rendered environment, so our findings support the sufficiency of a lightweight representation rather than a comparison between representations; whether selective use of 3D elements or gesture input adds to immersion or learning remains open. Beyond language learning, goal-conditioned situation generation with in-scene support and post-scene reflection may transfer to other domains of situated social interaction, such as clinical communication or conflict resolution. Domains that depend on physical interaction or conceptual reasoning would require different representations and generation strategies.

\subsection{Technical Limitations}
Generation latency, inconsistent imagery across scenes, and speech recognition errors interrupted immersion for some learners and likely affected narrative understanding. These are limitations of the current implementation rather than of the approach, and faster generation and consistency checks across generated components would address most of them. We also did not measure presence and narrative immersion against each other; examining how the two relate in generated practice is a direction for future work.

\section{Conclusion}

GELLI examined how interactive narratives generated at runtime could combine targeted language practice with learner authorship and conversational agency. Generated scenes provided opportunities for most assigned goals, and the controlled comparison found greater vocabulary recall and confidence than expert-authored videos targeting the same goals. Speaking performance improved in both conditions, with no significant between-condition difference. These single-session findings do not isolate individual components or establish retention or transfer.

Taken together, our findings suggest that practice can become an experience learners help create and learn through as it unfolds. Authoring let learners draw on remembered, anticipated, and imagined situations for preparation and play. Their accounts connected immersion to the relationship between these expectations, the generated representations, and responsive characters. Support made unfamiliar responses feel safer to attempt while preserving learners' responsibility for constructing what they said. Generation extended their initial intentions with language and developments they had not specified, making practice a way to explore possibilities, test expectations, and discover further things they wanted to learn.

Supporting this form of practice requires coordinating these elements as the experience develops. Authoring support must help learners convey and revise the situations they imagine; representations and character behavior must remain consistent while allowing believable surprises; and language support must enable experimentation without taking over. Narrative planning must also preserve reasons to use the target language, with room to relocate goals when they conflict with the unfolding interaction. GELLI contributes a design and empirical account of how learner authorship, generative narrative, and adaptive support can work together to make language practice both purposeful and open to discovery.

\bibliographystyle{ACM-Reference-Format}
\bibliography{sections/refs}

\end{document}